\documentclass[twocolumn,prl, aps,superscriptaddress,showpacs]{revtex4-2}
\usepackage{amsmath,amssymb}
\usepackage{bm}
\usepackage{color}
\usepackage{latexsym}
\usepackage{color}
\usepackage[english]{babel}
\usepackage{graphicx}
\usepackage{epsf}
\usepackage{subfigure}
\usepackage{amsfonts}
\usepackage{bm}

\newcommand{\slev}{5D_{5/2}}
\newcommand{\elev}{5P_{3/2}}
\newcommand{\glev}{5S_{1/2}}
\begin{document}

\title{Super-extended nanofiber-guided field for coherent interaction with hot atoms}
\author{R. Finkelstein}
\thanks{These authors contributed equally to this work}
\author{G. Winer}
\thanks{These authors contributed equally to this work}
\author{D.~Z. Koplovich} 
\author{O. Arenfrid}
\affiliation{Physics of Complex Systems, Weizmann Institute of Science, Rehovot 7610001, Israel}
\author{T. Hoinkes}
\affiliation{Vienna Center for Quantum Science and Technology,
TU Wien – Atominstitut, Stadionallee 2, 1020 Vienna, Austria}
\affiliation{Department of Physics, Humboldt-Universit{\"a}t zu Berlin, 10099 Berlin, Germany}
\author{G. Guendelman}
\author{M. Netser}
\affiliation{AMOS and Department of Chemical and Biological Physics, Weizmann Institute of Science, Rehovot 76100, Israel}
\author{E. Poem}
\affiliation{Physics of Complex Systems, Weizmann Institute of Science, Rehovot 7610001, Israel}
\author{A. Rauschenbeutel}
\affiliation{Vienna Center for Quantum Science and Technology,
TU Wien – Atominstitut, Stadionallee 2, 1020 Vienna, Austria}
\affiliation{Department of Physics, Humboldt-Universit{\"a}t zu Berlin, 10099 Berlin, Germany}
\author{B. Dayan}
\affiliation{AMOS and Department of Chemical and Biological Physics, Weizmann Institute of Science, Rehovot 76100, Israel}
\author{O. Firstenberg}
\affiliation{Physics of Complex Systems, Weizmann Institute of Science, Rehovot 7610001, Israel}

\begin{abstract}
We fabricate an extremely thin optical fiber that supports a super-extended mode with a diameter as large as 13 times the optical wavelength, residing almost entirely outside the fiber and guided over thousands of wavelengths (5 mm), in order to couple guided light to warm atomic vapor. This unique  configuration balances between strong confinement, as evident by saturation powers as low as tens of nW, and long interaction times with the thermal atoms, thereby enabling fast and coherent interactions.
We demonstrate narrow coherent resonances (tens of MHz) of electromagnetically induced transparency for signals at the single-photon level and long relaxation times (10 ns) of atoms excited by the guided mode.
The dimensions of the guided mode's evanescent field are compatible with the Rydberg blockade mechanism, making this platform particularly suitable for observing quantum non-linear optics phenomena.
\end{abstract}
%\setboolean{displaycopyright}{true}

\maketitle
\section{Introduction}

Efficient interaction between light and matter and particularly the faithful coherent mapping between photons and atomic excitations lie at the heart of many quantum optics processes and applications, such as quantum networks. One appealing platform is room-temperature atomic vapor, which is successfully employed in first generation quantum technologies, including atomic clocks and magnetometers \cite{Patton2014,Knappe2004} and quantum light sources and memories \cite{Lee2016,Finkelstein2018}.  
%Efficient interaction between atoms and light is essential in quantum technologies and in quantum information processing protocols. Room temperature alkali vapor provides a promising platform for such quantum technologies and is already utilized in first generation quantum sensors such as atomic clocks and magnetometers \cite{Patton2014,Knappe2004} [+add single photon sources?].
Light-matter interaction can be enhanced by a tight optical mode volume and by a collective coupling of this mode to an ensemble of atoms. While reduced mode volumes are achieved in small optical cavities \cite{Alaeian2020} or with tightly focused beams in free space \cite{Maser2016,Jia2018a}, they are typically incompatible with large ensembles of atoms due to the associated short Rayleigh range $z_\mathrm{R}$.
\begin{figure}[ht!]
    \centering
    \includegraphics[width=\columnwidth]{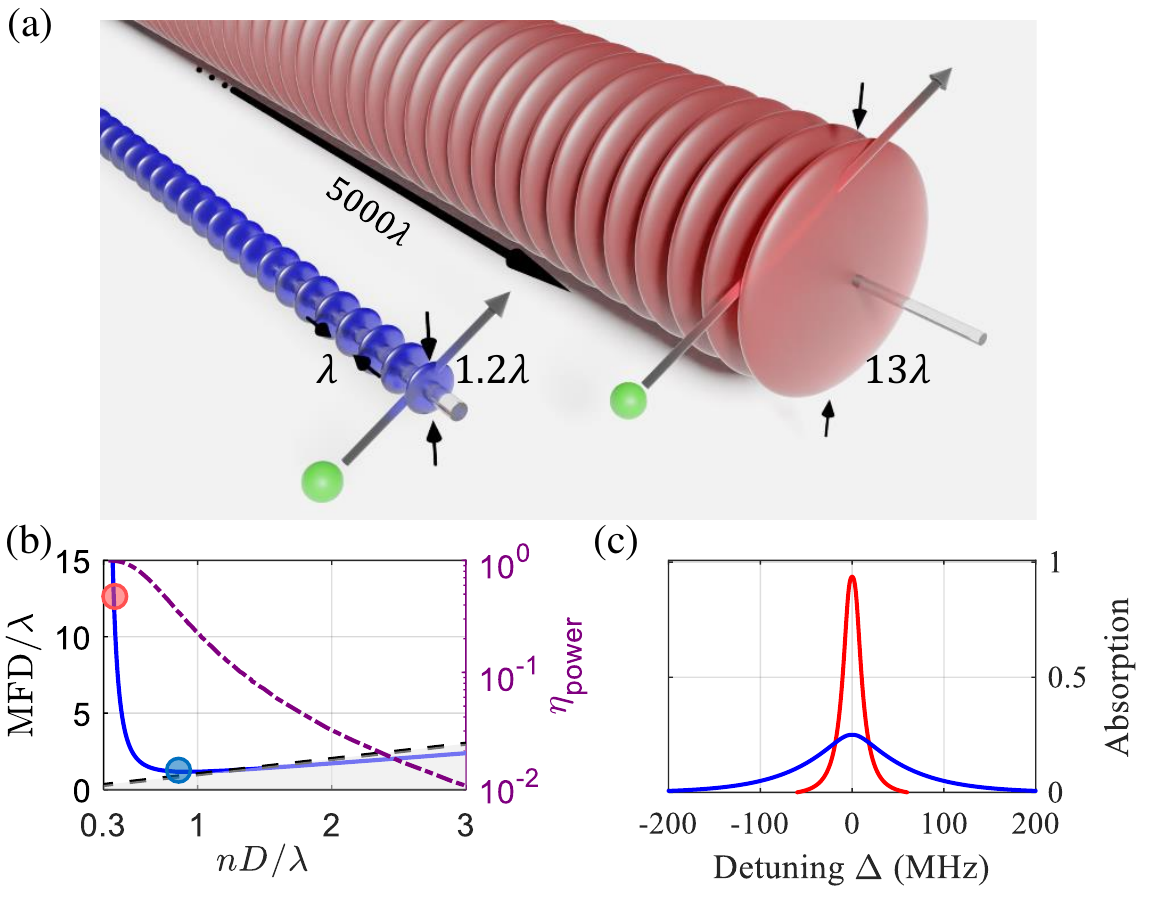}
        \caption{\textbf{A super-extended evanescent field of a nanofiber interacting with atomic vapor.} (a) Illustration of an optical mode surrounding a thin optical fiber. The fiber on the left (fiber diameter $D=0.9\lambda/n$) has a guided mode with field diameter MFD=$1.2\lambda$, comparable with the wavelength $\lambda$. The fiber on the right ($D=0.37\lambda/n$) has a mode extending to MFD=13$\lambda$, guided over a distance of $5000\lambda$. (b) Mode field diameter (in units of $\lambda$) as a function of fiber diameter (in units of wavelength in matter $\lambda/n$). Blue and red circles mark the parameters of the two fibers shown in (a). Dashed black line marks the physical fiber dimensions; the MFD diverges from this line around $nD/\lambda < 1$. The dashed purple line shows the fraction of the power residing outside the fiber $\eta_\mathrm{power}$ and is larger than $99\%$ for the thin fiber. (c) Calculated Doppler-free absorption spectra for Rb vapor in the evanescent field of a $D=0.9\lambda/n=500$~nm (blue) and $D=0.37\lambda/n=200$~nm (red) silica fibers ($n=1.45$). The effect of transit-time broadening is clearly apparent. 
        }\label{fig:Figure1}
\end{figure}

An alternative approach employs a tightly confined optical mode that is supported by a low-loss waveguide over an extended length  \cite{Spillane2008,Stern2013,Skljarow2020a,Nayak2007,Sague2007,Bajcsy2009}. 
% Evanescent field: a tight mode creates TOF and limits atom number
The optical mode guided by a dielectric structure with dimensions below the optical wavelength $\lambda$ extends beyond the structure boundaries as an evanescent field and can interact with the surrounding atomic ensemble. In this case, the field typically decays over a range of $\sim\lambda$ away from the structure \cite{Stern2013,Hendrickson2010,Ritter2015}. Such tight mode confinement is advantageous for processes requiring high intensity or a steep field gradient, \emph{e.g.}, non-linear optics at ultra-low power levels or atomic trapping \cite{Foster2004,Vetsch2010}. However, for coherent light-matter interaction with thermal atomic ensembles, tight confinement may limit the effectiveness of the interaction. Transverse motion of the atoms through the optical mode results in transit-time broadening and in a reduction of the absorption cross-section \cite{Stern2013,Hendrickson2010,Jones2015,Skljarow2020a}. \pagebreak 
\\
\\
   Additionally, the small mode volume limits the fraction of atoms simultaneously interacting with the mode, and the proximity of these atoms to the dielectric waveguide may lead to atom-surface interactions that impair the coherent interaction with the guided field \cite{Epple2014a,Sague2007}. 
%Finally, the absorption cross-section is also fundamentally limited by the fraction of the optical field which resides outside the core material and has a probability to interact with the atoms. 
Finally, for a given uniform atomic density, the collective light-matter coupling strength (or the optical depth) is limited by the fraction of the optical field which resides outside the core material.
%Finally, deterministic, strong, photon-photon interactions are limited by the fraction of the optical field which resided outside the core material.

%nano fiber: what they are and who uses them

In this work, we tackle these challenges by realizing a guided optical mode with an evanescent field part that extends several wavelengths away from the waveguide surface, as illustrated in Fig.~\ref{fig:Figure1}(a). This unique mode is supported by the extremely thin waist of a tapered optical fiber. Tapered optical fibers, with waist diameters as small as a few hundred nanometers, have been shown to enable single mode operation with high transmission \cite{Tong2003}. In the past decade, these nanofibers were used in a multitude of applications from atom trapping \cite{Vetsch2010}, through sensing \cite{Korposh2019}, to chiral quantum optics \cite{Petersen2014} and cavity QED \cite{Takao2015,Bechler2018}. Previous demonstrations of an interface between thermal vapor and a nanofiber have shown several appealing features, such as polarization rotation, electromagnetically induced transparency \cite{Jones2015}, and non-linear effects at low power \cite{Hendrickson2010}. However, all of these have been limited by transit time broadening, as well as by the interaction between the vapor and the waveguide surface.

Here we show that reducing the diameter of a nanofiber to around $0.37\lambda/n$ (where $n$ is the refractive index of the fiber core) yields a super-extended evanescent mode, and that this mode greatly suppresses the above limitations. In our system, the mode extends to a diameter of $2w_0 \approx13\lambda$ and is guided over 5 mm (50 times larger than $z_\mathrm{R}= \pi w_0^2/\lambda$), with over $99\%$ of the optical power residing outside the core material. We interface this fiber with atomic vapor and perform one- and two-photon spectroscopy, as well as saturation and temporal transient measurements. The unique characteristics of the system provide for spectral features much narrower than previously measured and correspondingly longer coherence times, thus establishing its potential for more intricate processes and opening a path to coherent light-matter interactions with thermal vapor in evanescent fields.

\section{Super-extended evanescent field}
A guided optical mode can be characterized by the mode field diameter (MFD), defining an area containing $(1-e^{-2})$ of the optical power. For vacuum-clad fibers, the MFD of the fundamental mode $\mathrm{HE_{11}}$ depends on a single parameter: the fiber diameter $D$ in units of wavelength in matter $\lambda/n$. We show calculations of the MFD as a function of this parameter in Fig.~\ref{fig:Figure1}(b). As the fiber diameter is decreased, the MFD initially follows the diameter of the fiber core. In this regime, a significant fraction of the energy is contained inside the fiber core and its remainder resides in an evanescent field, decaying over $\sim\lambda$. When the fiber diameter is further narrowed below $nD/\lambda = 1$, it can no longer support a mode residing predominantly in the core. Yet a single bound solution, with most of the energy residing outside the core, always exists \cite{Snyder1983}.

In this regime, the MFD varies steeply with the fiber diameter. This is demonstrated by comparing two cases, illustrated in Fig.~\ref{fig:Figure1}(a) and marked by circles in Fig.~\ref{fig:Figure1}(b). A fiber with $D=0.9\lambda/n$ has MFD = 1.2$\lambda$ (blue circle), whereas a narrower fiber with $D=0.37\lambda/n$ has MFD = 13$\lambda$ (red circle). The latter, which we realize in this work, is a super-extended mode that contains a remarkably high fraction of $>99\%$ of the power in the vacuum cladding. This mode is well approximated by a modified Bessel function \mbox{$\mathcal{E}(r>D/2)=\mathcal{E}_0 K_0(\kappa r)$} \cite{Tong2004}, where \mbox{$\kappa=\sqrt{\beta^2-(2\pi/\lambda)^2}$} is the transverse component of the wavevector  ($\beta$ is the propagation constant, \emph{i.e.}, the longitudinal component of the wavevector). For $\lambda=780$ nm, a $D=200$ nm silica fiber can guide a mode with MFD = 10 $\mu$m ($\kappa^{-1}\simeq 0.44\mathrm{MFD}=4.43~\mu\mathrm{m}$), such that the mode area is 2500-times larger than the fiber cross-section. The ability to produce such a waveguide by tapering down a standard optical fiber, with adiabatic following of the fundamental mode, is an exciting capability which lies at the edge of feasibility for waveguide tapering \cite{Stiebeiner:10,Sumetsky2006}.

Both fibers highlighted in Fig.~\ref{fig:Figure1}(a) guide the optical modes over many times the equivalent $z_\mathrm{R}$ in free space. However, when interfaced with atomic vapor, the interaction characteristics will strongly depend on mode confinement. Thermal ballistic motion, with thermal velocity $v_\mathrm{T}$ along both the longitudinal and the transverse wavevectors, leads to motional broadening. For modes extending to $\kappa^{-1}{\sim}\lambda/2$ away from the fiber, the transverse transit-time broadening $\Gamma_\mathrm{tt}=\sqrt{2}\kappa v_\mathrm{T}$ (full width at $1/e$) \cite{Hall} becomes similar to the longitudinal Doppler broadening $\sigma=\beta v_\mathrm{T}$. In contrast, for the super-extended optical mode with $\kappa^{-1}\gtrsim 5\lambda$, the transverse motional broadening is suppressed by more than one order of magnitude.
%Due to their thermal ballistic motion, atoms transit through the tight mode around the thicker fiber [left in Fig.~\ref{fig:Figure1}(a)] quicker than through the mode guided by the thin fiber [right in Fig.~ \ref{fig:Figure1}(a)]. This leads to transit-time broadening or decoherence, which

We note that while many Doppler-free techniques exist, including compensation by light shift \cite{Lahad2019,Finkelstein2019}, transit-time broadening cannot be easily mitigated by purely optical means. The calculated (Doppler-free) absorption spectra for both fibers presented in Fig.~\ref{fig:Figure1}(c) indeed show a 10-fold decrease in spectral width for the thinner fiber and a corresponding enhancement in atomic absorption cross-section. In this calculation, we average over atoms with different thermal velocities traversing the two-dimensional field distribution and neglect atomic trajectories which hit the nanofiber. This is a valid approximation for the super-extended mode due to the large ratio of mode-field area to fiber cross-section.

\section{Experimental system}

\begin{figure} [tb]
    \centering
    \includegraphics[width=\columnwidth]{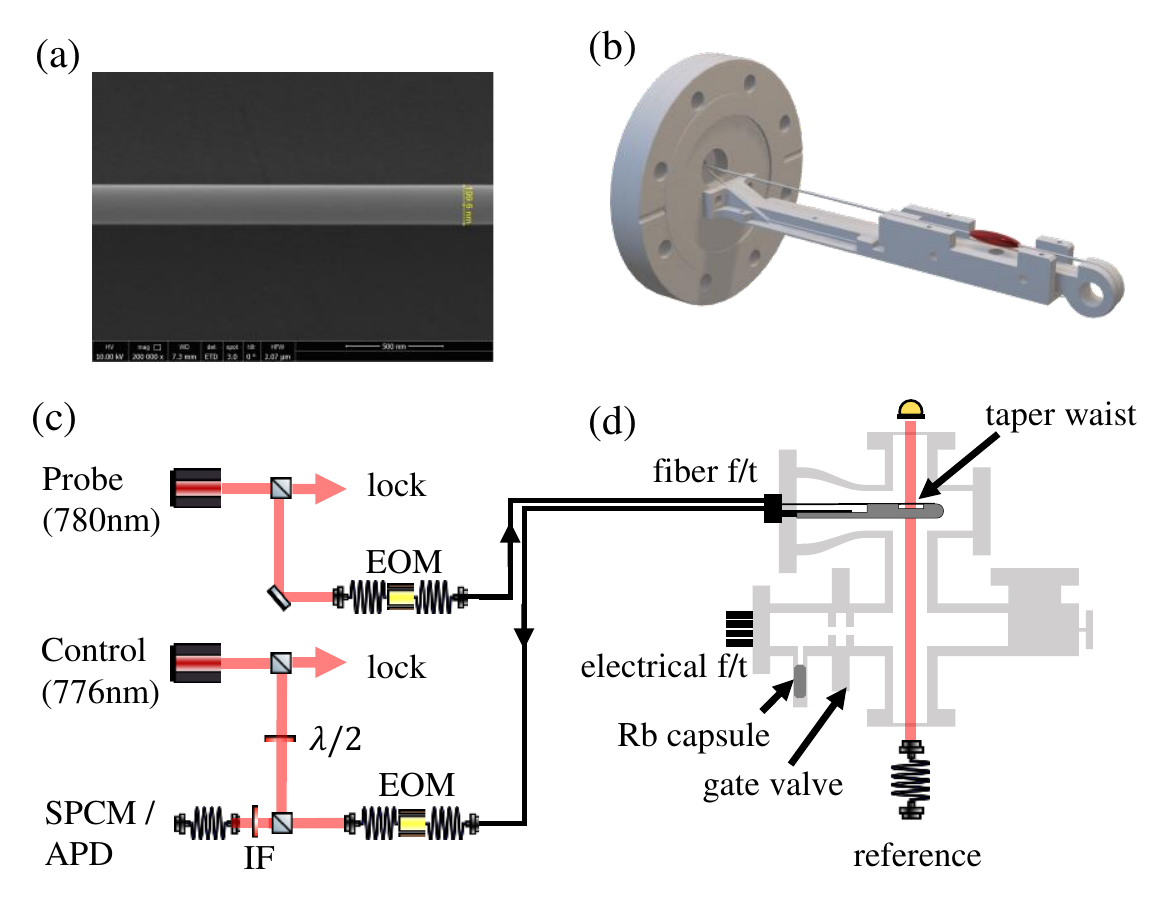}
    % The 780 nm probe light is coupled to a single mode fiber which is spliced to the nano fiber. The pre-taper SM fiber polarization can be controlled via a fiber stressor. The 776 nm control counter propagates w.r.t the probe. Its intensity is modulated by a fiber coupled PPLN EOM. Probe transmission measurement is done with a SPCM.
    % The ladder excitation scheme involves the D2 transition of $^{85}Rb$, coupling $5S_{1/2} F=3$ to $5P_{3/2} F=2$, and the upper state $5P_{3/2}$.
    \caption{\textbf{Experimental system.} (a) Scanning electron microscope image of a tapered fiber waist, with a nominal diameter of 200 nm. (b) Custom, vacuum-compatible, fiber mount with external optical access and a UHV fiber feedthrough (f/t).  (c) Schematic of the optical setup. Probe and control fields are coupled into the tapered fiber in counter-propagating directions. Fiber-coupled electro-optical modulators (EOMs) shape the temporal intensity of the two fields. Polarizing beam-splitter (PBS) picks out the outgoing probe, which is further filtered by a band-pass interference filter (IF) and sent to a single-photon counting module (SPCM) or to an avalanche photodiode (APD), allowing us to monitor down to pW powers. (d) Vacuum chamber houses the fiber and a natural abundance metallic Rb pellet. A free-space path provides an absorption reference.}
    \label{fig:Vacuum setup}
\end{figure}

The experimental setup is presented in Fig.~\ref{fig:Vacuum setup}. A fiber with super-extended evanescent field is fabricated by tapering a silica fiber down to a nominal diameter $D$=200 nm (with variations better than $\pm5\%$) in a heat and pull method \cite{Stiebeiner2010}. Fig.~\ref{fig:Vacuum setup}(a) shows a scanning electron microscope image of the nanofiber waist of a tapered fiber that we have pulled using the same parameters and flame trajectory as the tapered fiber used in the subsequent experiments. To fulfill the adiabaticity criterion, the fiber is tapered down from a diameter of 125 $\mu$m to the final diameter of 200 nm over a length of 3.3 cm. The fiber is then glued to a custom mount [Fig.~\ref{fig:Vacuum setup}(b)], which provides three axes of optical access and an ultra-high vacuum-compatible fiber feedthrough. The mount is installed in a vacuum chamber [Fig.~\ref{fig:Vacuum setup}(d)], which is wrapped in resistive heating elements and surrounded by a thermally insulating polyurethane enclosure, allowing to stabilize the system temperature. Rubidium vapor is released into the chamber by breaking a glass capsule containing a metallic rubidium pellet (at natural abundance). We set two different temperature regions in the chamber; the area containing the capsule is kept cooler such that it remains a rubidium reservoir, and its temperature sets the vapor pressure inside the cell.

%Fig.~\ref{fig:Vacuum setup}

\begin{figure}[tb]
    \centering
    \includegraphics[width=\columnwidth]{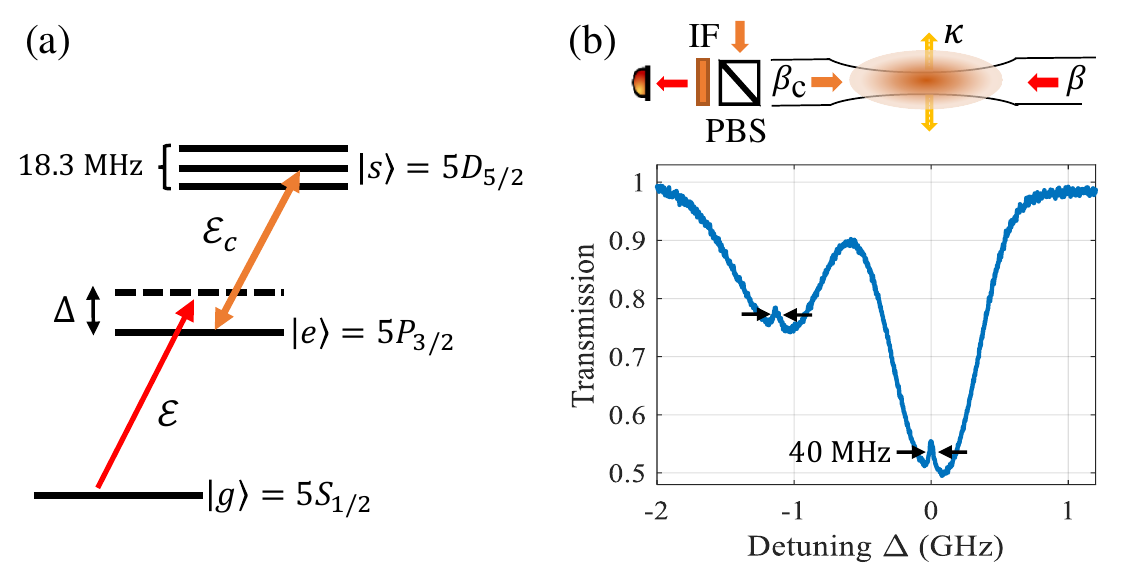}
    \caption{\textbf{Overview of the experiment.} (a) Level structure of rubidium atoms used in the experiment. The ground and intermediate levels are coupled by a probe field $\mathcal{E}$, detuned by $\Delta$ from the resonance frequency. The intermediate level can be coupled to a doubly excited level $\slev$ by a control field $\mathcal{E}_\mathrm{c}$. The $\slev$ level in $^{85}$Rb is composed of several tightly spaced hyperfine states. The three states accessible in our experiment are spaced over 18.3 MHz.  
    %The homogeneous decoherence rates of the excited levels are denoted by $\gamma$ and $\gamma_\mathrm{gs}$.
    (b) Typical transmission spectrum of Rb vapor coupled to a super-extended evanescent field guided by a tapered optical fiber. The probe and control fields with propagation constants $\beta$ and $\beta_\mathrm{c}$ counter-propagate in the fiber. The two-photon transition is nearly Doppler-free as the propagation constants differ by about 0.5\%. Both evanescent fields have a similar decay length $\kappa^{-1}$ in the transverse direction. The control field is filtered by polarization and narrow-band interference filter.}
    \label{fig:Experimental config}
\end{figure}

We utilize the electronic ladder transitions of rubidium shown in Fig.~\ref{fig:Experimental config}(a). A probe field at 780 nm probes and drives the D2 transition $\glev \rightarrow \elev$ in all experiments. Two-photon spectra, demonstrated in Fig.~\ref{fig:Experimental config}(b), are obtained by adding a control field at 776 nm that drives the transition $\elev \rightarrow \slev$. The control and probe lasers are frequency stabilized and fed into the nanofiber in counter-propagating directions. We show the optical setup in Fig.~\ref{fig:Vacuum setup}(c). The vacuum chamber also has a long free-space path, shown in Fig.~\ref{fig:Vacuum setup}(d), for measuring a reference absorption spectrum. More details on the fiber and setup are given in Methods. 

The transmission of the fiber at 780 nm is $\sim90\%$ after fabrication, and it drops to $\sim 70\%$ after international transfer, splicing to commercial fibers, and installation in the chamber. However, the transmission may further degrade over time due to rubidium adsorption on the fiber surface. We are able to mitigate this effect and prevent the transmission degradation by heating up the fiber using $1~\mu$W of light at 776 nm that is kept constantly on.

\section{Super-extended evanescent field interfaced with thermal vapor}
% {\blue{I_sat Experiment.}}

\begin{figure}[tb]
    \centering
    \includegraphics[width=\columnwidth]{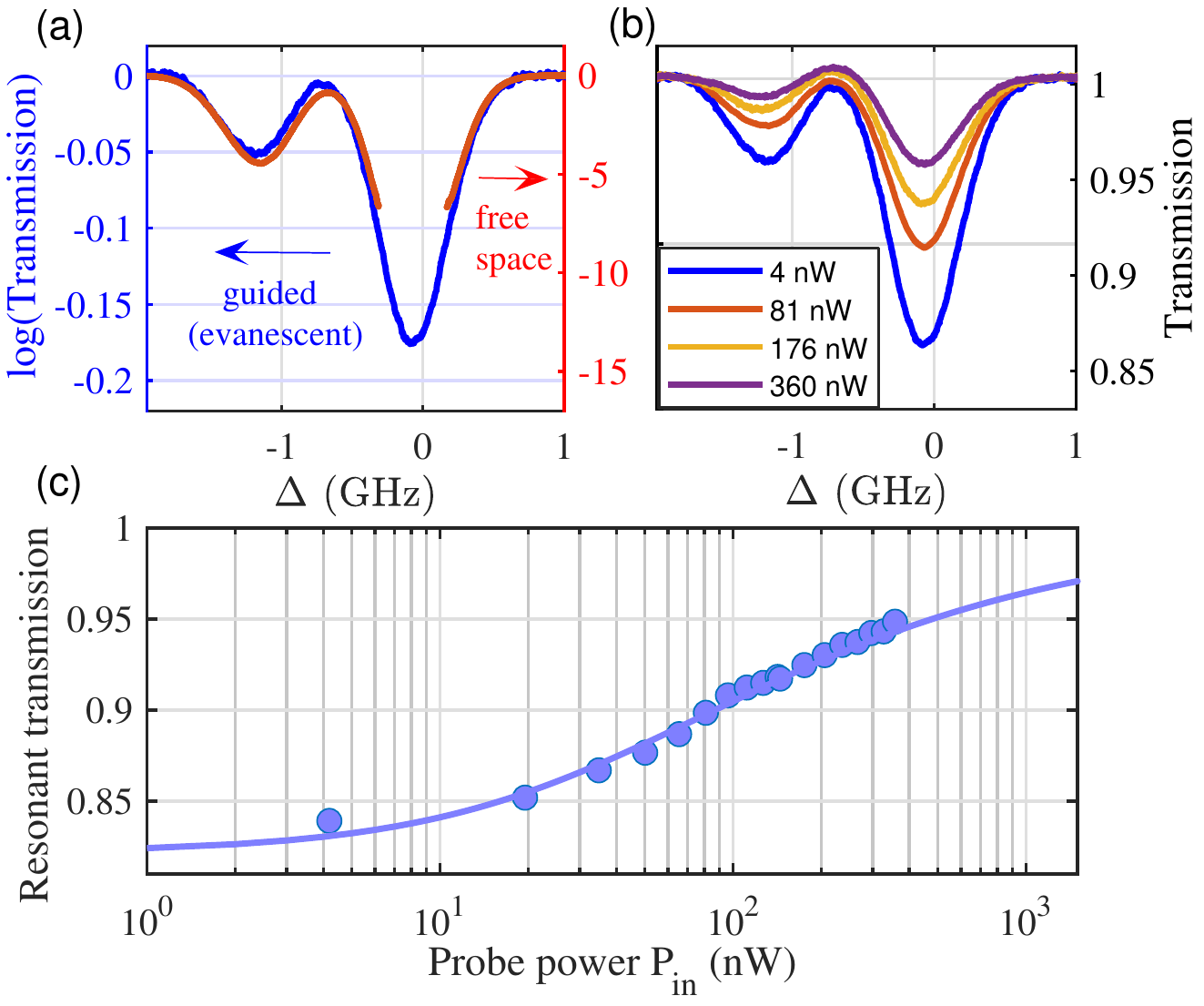}
    \caption{\textbf{Transmission spectra and low-power saturation.} (a) Transmission spectrum of light guided in a tapered fiber with super-extended evanescent field surrounded by Rb vapor with natural abundance (blue). Two distinct dips correspond to two Rb isotopes. A free-space absorption spectrum is measured simultaneously as a reference, through a 280-mm-long optical path across the chamber (red; transmission below $10^{-3}$ around the $^{85}\mathrm{Rb}$ dip is governed by noise and not shown).  (b) Transmission spectra for different probe powers $P_{\mathrm{in}}$. (c) Resonant transmission as a function of $P_{\mathrm{in}}$. Solid line is a fit to a saturation model. All data in (a)-(c) are normalized by the off-resonance transmission.}  %Inset: 2 GHz probe detuning scans, showing absorption dips of both Rb isotopes, taken at 4 probe power levels.}
    \label{fig:Isat}
\end{figure}

We begin with characterizing the super-evanescent mode by measuring the atomic absorption spectra with only the probe light present. In Fig.~\ref{fig:Isat}(a), we show the transmission spectrum of light guided by the tapered fiber and compare it with the free-space transmission spectrum of a large diameter (2 mm) beam measured simultaneously. The spectrum exhibits two transmission dips corresponding to $^{87}$Rb and $^{85}$Rb isotopes according to their natural abundance. The width of these absorption lines is dominated by Doppler broadening due to longitudinal atomic motion along the fiber axis. 
Importantly, we do not observe any additional broadening due to the transverse motion across the evanescent mode,
when comparing the spectrum (blue line) to the free-space spectrum (dashed red line) and when fitting it to a numerical model (not shown) that accounts for the multi-level structure of rubidium. This is already an indication of an extended mode spanning at least several wavelengths.

In Fig.~\ref{fig:Isat}(b), we plot absorption spectra for different probe powers. As the probe power increases, the absorption decreases, with the resonant transmission approaching unity already at a few hundreds of nW. We summarize these results in Fig.~\ref{fig:Isat}(c) by plotting the transmission at the $^{85}$Rb $F{=}3{\rightarrow}F'{=}4$ resonance for different probe powers and fitting the data to a saturation model for an inhomogeneously broadened ensemble of two-level systems \cite{Akulin1992} $T=\mathrm{exp}(-\mathrm{OD_0}/\sqrt{1+P_\mathrm{in}/P_\mathrm{sat}})$, where $\mathrm{OD}_0$ is the resonant, Doppler-broadened, optical depth of the ensemble, $P_\mathrm{in}$ is the probe power, and  $P_\mathrm{sat}$ defines the saturation power. We find that the saturation power is $P_\mathrm{sat}=35\pm7$ nW. Given the imperfect transmission through the bare tapered fiber, we estimate the actual saturation power (reaching the atoms) to be $29\pm6$ nW. We note that for an extended evanescent field with decay length $\kappa^{-1}$, the saturation parameter per atom (often denoted as $s$) increases as the atom-nanofiber distance decreases and, for $P_\mathrm{in}=P_\mathrm{sat}$,  becomes larger than unity for atoms within a distance of $\kappa^{-1}$ from the nanofiber axis.

This low saturation power is characteristic of a tightly confined optical mode. Even though the realized mode is larger than typical evanescent fields, the measured saturation power is identical, up to experimental uncertainty, to saturation powers reported across several platforms interfaced with hot vapor \cite{Stern2013,Hendrickson2010,Spillane2008}. This universality stems from the trade-off which sets the saturation threshold: Rabi frequency and decoherence rate (dominated by transit-time broadening) both increase linearly with 1/MFD.  
As a consequence, we find that this type of non-linearity does not benefit from tighter mode confinement and, in fact, super-extended evanescent fields are advantageous as they enable the same saturation level with better coherence times.    
On the flip side, this trade-off also implies that $P_\mathrm{sat}$ measurements cannot unambiguously confirm the dimensions of the optical mode.
%*** For large modes, typically in free-space beams, the mode diameter can be directly extracted by dividing the saturation power by the theoretical saturation intensity of the relevant transition. However, in tight modes the decoherence rate $\gamma$ is dominated by transit time broadening which scales inversely with the mode diameter. This is in contrast to the local intensity which scales quadratically with the MFD. For this reason, the saturation power we measure is identical to systems with much tighter modes \cite{Stern2013,Spillane2008,Hendrickson2009}, and the transmission saturation measurements can not unambiguously confirm the dimensions of the optical mode. ***

% {\blue{Transient Experiment.}}

In order to determine the transit-time broadening and thus the dimensions of the mode, we perform temporal transient measurements. In essence, we prepare a saturated atomic population and measure the rate at which the excited atoms leave the interaction region. In practice, we monitor the response of the resonant transmission to a sudden change in probe power in a pump-probe-like experiment. 

The dynamics is governed by the optical Bloch equations for the excited population $\rho_\mathrm{ee}$ and the atomic coherence (optical dipole) $\rho_\mathrm{eg}$ 
\begin{equation}
    \partial_t \rho_\mathrm{ee}=-\Gamma\rho_\mathrm{ee}+2\mathrm{Im}(\Omega^*\rho_\mathrm{eg}),
\end{equation}
\begin{equation}
    \partial_t \rho_\mathrm{eg}=-
    \sigma\rho_\mathrm{eg}+i\Omega(1-2\rho_\mathrm{ee}),
\end{equation}
where $\Gamma$ is the depopulation rate, \emph{i.e} the population decay rate due to both spontaneous emission and transit time across the evanescent field, and $\sigma$ is the total effective decoherence rate, dominated by Doppler dephasing $\sigma\approx \beta v_\mathrm{T}$ ($v_\mathrm{T}=180$ m/s). The probe Rabi frequency is given by $\Omega=G\mathcal{E}$, where $G$ is the ensemble-field coupling constant, and $\mathcal{E}$ is the probe electric field.
In turn, the propagation of the probe field is governed by the equation of motion for the slowly varying envelope of the electric field  $(c\partial_z+\partial_t){\mathcal{E}}=iG\rho_\mathrm{eg}$ \cite{Fleischhauer2000}, which reduces to 
\begin{equation}
    \mathcal{E}_\mathrm{out}=\mathcal{E}_\mathrm{in}+i\rho_\mathrm{eg}GL/c
\end{equation}
for an optically thin medium. Here the outgoing field $\mathcal{E}_\mathrm{out}$ is given by the sum of the incoming field $\mathcal{E}_\mathrm{in}$ and the field scattered by the atomic dipoles. We can therefore extract the evolution of the atomic coherence by monitoring the difference \mbox{$\Delta\mathcal{E}=(P_\mathrm{in}-P_\mathrm{out})/\sqrt{P_\mathrm{in}}$}, where $P_\mathrm{out}$ and $P_\mathrm{in}$ are measured powers on and off resonance, respectively. Such analysis is akin to that employed in cavity ring-up spectroscopy \cite{Rosenblum2015}. 

\begin{figure}[tb]
    \centering
    \includegraphics[width=\columnwidth]{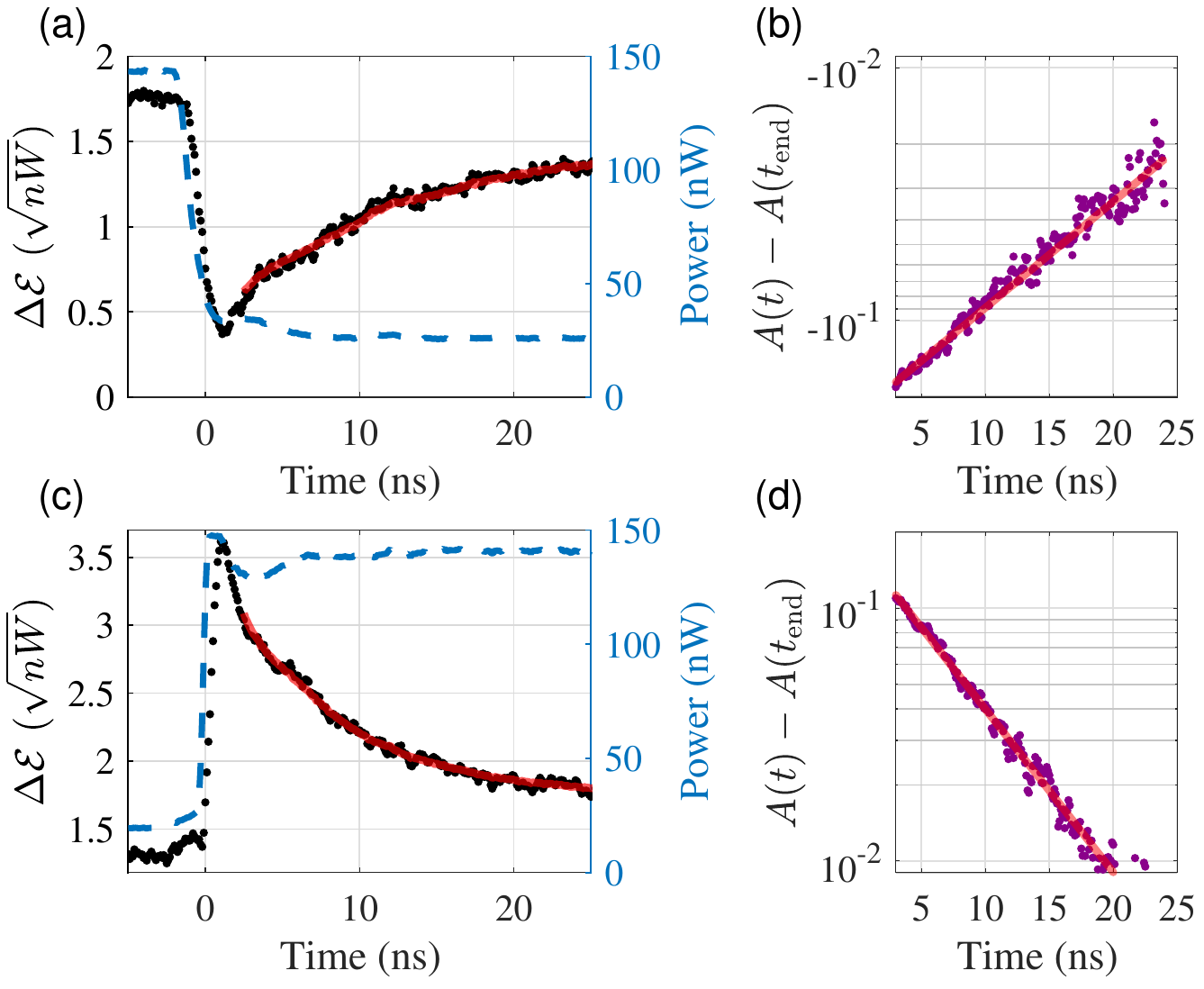}
    \caption{\textbf{Temporal dynamics in the nanofiber-vapor system.} (a) The transient response of the atomic coherence (optical dipole) $\rho_\mathrm{eg}$, quantified by the difference $\Delta\mathcal{E}$ between the incoming and outgoing probe (black dots, see text) after an abrupt reduction of the probe power (dashed blue line). The atomic coherence initially follows the rapidly changing incoming field and then slowly relaxes due to motional exchange of pumped and unpumped atoms and due to radiative decay. (b) Absorption $A(t)$ following the abrupt power reduction, plotted as the difference from the final steady-state value $A(t_\mathrm{end})$ in a semilog scale. Solid line is a fit to a pure exponential decay, also plotted in (a), from which the relaxation time $\tau_\mathrm{fall}$ is extracted. (c) An abrupt increase in the laser power again initiates a fast following of the field and then a slower decay, determined by the optical pumping rate. (d) Absorption following the abrupt power increase, providing the faster relaxation time $\tau_\mathrm{rise}$ from which the probe Rabi frequency can be determined.}
    \label{fig:transient}
\end{figure}

Figure \ref{fig:transient} shows the results of such an experiment. We start with a probe power of $\sim 4P_\mathrm{sat}$, \emph{i.e.,} above the saturation power, and abruptly attenuate it to $0.6P_\mathrm{sat}$. 
Equations (1) and (2) under the condition $\sigma \gg \Gamma,\Omega$ result in an over-damped solution with a two-step decay process: a fast decay with rate $\sigma$ followed by a slower decay with rate $\Gamma$. 
Indeed we observe in Fig.~\ref{fig:transient}(a) an initial short transient, where the atomic coherence follows the fast change in the incoming field, and subsequently a slow relaxation due to equilibration of atomic population.

To quantify the slow relaxation rate, we plot in Fig.~\ref{fig:transient}(b) the absorption $A(t)=1-P_\mathrm{out}(t)/P_\mathrm{in}(t)$. The (pumped) excited atoms are gradually replaced by (fresh) ground-state atoms that enter the interaction region, and correspondingly the medium relaxes from low to high absorption. We measure an exponential relaxation time of $
\tau_{\mathrm{fall}}=10.7 \pm 0.5 $ ns, corresponding to a depopulation rate of $\Gamma=14.9 \pm 0.8$ MHz (hereafter $1~\mathrm{MHz}\equiv 10^6\cdot 2\pi~\mathrm{rad}/ \mathrm{s}$). After subtracting the radiative decay rate (6 MHz), we attribute the remaining rate of $\Gamma_\mathrm{tt}=9$ MHz to transit time broadening.  This is consistent with the numerical calculation of transit time broadening of $\Gamma_\mathrm{tt}=\sqrt{2}\kappa v_\mathrm{T}=9$ MHz for a fiber with diameter $D=200 $ nm ($\kappa=1/4.428~ \mu m^{-1}$), as plotted in Fig.~\ref{fig:Figure1}(c). %We note that the decoherence rate due to transit time alone is half of the depopulation rate, \emph{i.e.}, 4.5 MHz.
%We note that the pure exponential relaxation rate due to transit time is unique to evanescent fields that decay exponentially in the radial direction and results in an approximately Lorentzian line-shape. This is in contrast with Gaussian beams which yield a cusp-like line-shape $\exp(-|\Delta|w_0/v_\mathrm{T})$ \cite{Biraben1979}. 

%The pumped atomic medium relaxes from low absorption (corresponding to an atomic population pumped to the excited state) to high absorption (corresponding to a fresh population at the ground state). 
%This rapid change toggles atomic absorption between different steady state values which correspond to values measured in Fig.~\ref{fig:Isat}. 

% ring up measurement
Upon ramping the strong field back on [Fig. \ref{fig:transient}(c)], we observe again a fast transient response followed by an exponential relaxation, with an absorption decay time [Fig. \ref{fig:transient}(d)] of $\tau_\mathrm{rise}=6.7 \pm 0.3$ ns. We attribute this time scale to optical pumping to the excited state at a rate of \mbox{$\Gamma+4\Omega^2/\sigma=24 $ MHz}, from which we infer a conversion of optical power to Rabi frequency of $\Omega(P_\mathrm{in}) = 2~\mathrm{MHz} \sqrt{P_\mathrm{in}[\mathrm{nW}]}$. This conversion ratio is in good agreement with a weighted average over the calculated field distribution and, specifically, it characterizes atoms at a distance of 2 $\mu\mathrm{m}$ ($\sim0.5\kappa^{-1}$) from the fiber.  
%is in good agreement with the estimated MFD of 10 $\mu$m. 

%Applying this to the measured saturation power, we find $\Omega(P_\mathrm{sat})=11\pm1 $ MHz. This value corresponds to a decoherence rate $\sqrt{2}\Omega(P_\mathrm{sat})=15.5$ MHz, which is slightly higher than $\Gamma_\mathrm{tt}=9$ MHz, indicating an additional pure dephasing process which is on the same order of magnitude as the decoherence rate due to loss of population of atoms exiting the interaction region. 

% EIT
\begin{figure}[tb]
    \centering
    \includegraphics[width=\columnwidth]{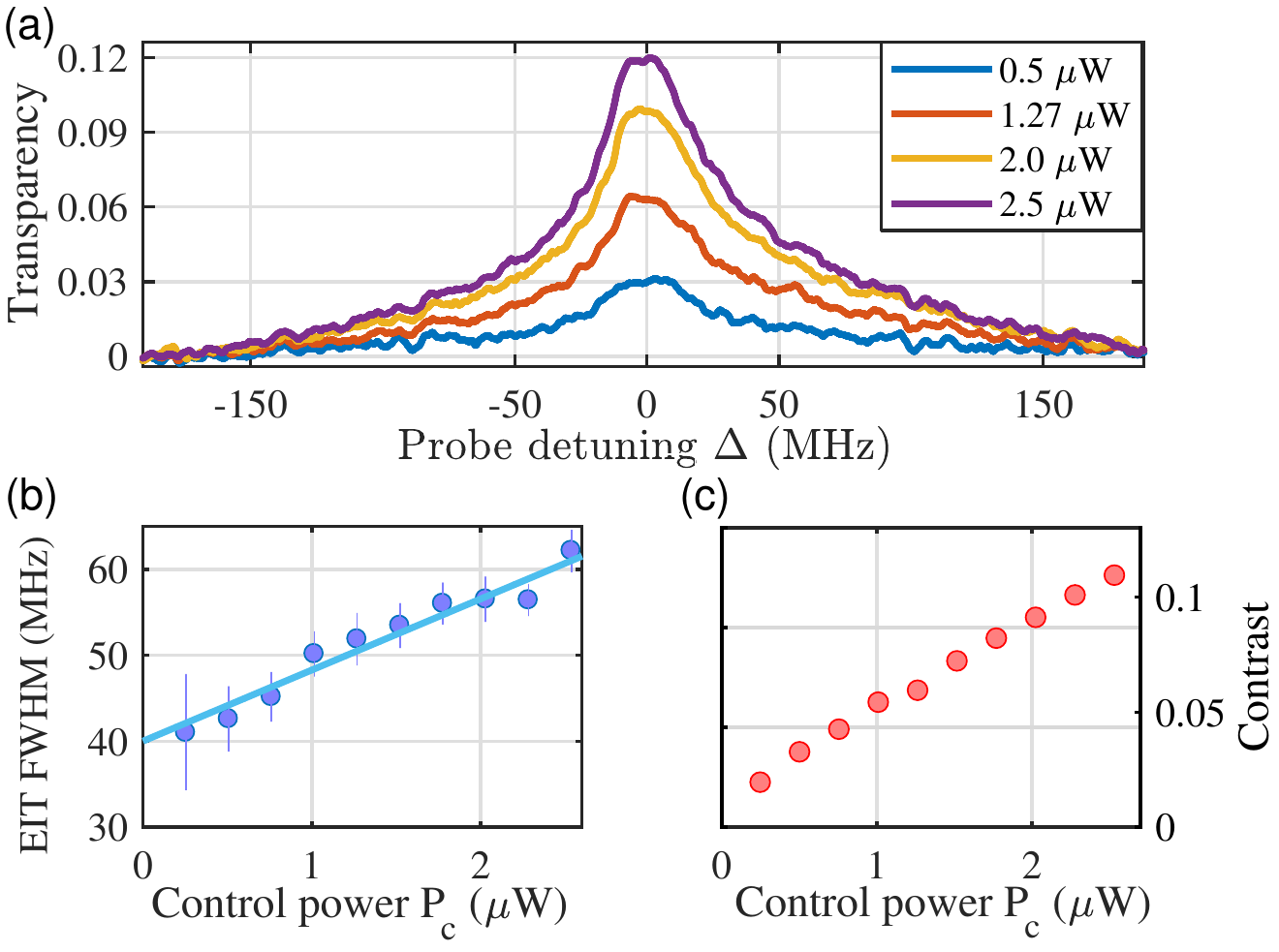}
    \caption{\textbf{EIT in a bi-chromatic super-extended evanescent field.} (a) Transmission spectrum of a weak probe in the presence of a control field of varying  power $P_\mathrm{c}$, counter-propagating in the nanofiber. (b) Full width at half maximum (FWHM) of the transparency resonance. Solid line is a fit to a power-broadening model. (c) Contrast of the transparency resonance.}
    \label{fig:figEIT}
\end{figure}

We now move on to study a coherent, two-photon process, taking advantage of the long coherence time provided by the super-extended optical mode. We use a counter-propagating configuration that is nearly Doppler free, as the relative mismatch of the propagation constants $(\beta_\mathrm{c}-\beta)/\beta$ is only about 0.5$\%$. This reduces the longitudinal Doppler broadening to 0.5$\%$ of its value for the one-photon transition. We add the counter-propagating control field and observe the appearance of narrow electromagnetically-induced transparency (EIT) peaks inside the one-photon absorption lines, as shown in Fig.~\ref{fig:Experimental config}(b). Notably, most two-photon processes observed in atomic vapor via a waveguide have been limited to cascaded absorption \cite{Stern2013,Hendrickson2010,Skljarow2020a}. In our system, the small transit-time broadening and high signal-to-noise ratio enable the coherent effect of induced transparency, originating from interference of two possible excitation pathways. Furthermore, we are able to measure EIT with $P_\mathrm{in}\simeq$ pW or, equivalently, with probe photons entering the medium at a rate of few photons per $\mu s$ such that the atoms interact with less than a single probe photon (on average) during their lifetime in the mode. This is a pre-requisite for observing quantum non-linear optics.

In our configuration, the control field is also guided by the fiber with a similar, super-extended, mode and not applied externally. To leading order, the coherent two-photon process depends on the product of the probe and control fields $\mathcal{E}(r)\cdot\mathcal{E}^*_\mathrm{c}(r)$, which varies with time while a given atom traverses the mode. Both evanescent fields are of the form $K_0(\kappa r)\approx e^{-\kappa r}/\sqrt{r}$, such that their product is approximately proportional to $e^{-2\kappa r}/r$, yielding an increased two-photon transit time broadening of $2\times\Gamma_\mathrm{tt}$. %(note that higher order contributions, e.g. power broadening $\propoto |Omega_c^2|$ will be even worse...)

In Fig.~\ref{fig:figEIT}(a), we plot the transmission around the EIT resonance for different control powers, normalized by the one-photon absorption (in the absence of a control field). We note that increasing the probe power above a few pW reduces the EIT contrast, and, for probe powers higher than $P_\mathrm{sat}$, we have observed cascaded absorption. 

We present in Fig.~\ref{fig:figEIT}(b,c) the full width at half maximum (FWHM) of the EIT resonance $\Gamma_\mathrm{EIT}$ and the EIT contrast for different control powers $P_\mathrm{c}$. The width and contrast increase linearly with $P_\mathrm{c}$. The solid line in Fig.~\ref{fig:figEIT}(b) is a fit to an EIT power-broadening model of the form $\Gamma_\mathrm{EIT} = \alpha P_\mathrm{c}+\gamma$. By extrapolating to $P_\mathrm{c}=0$, we find the effective width of the $\glev\rightarrow\slev$ transition $\gamma=40\pm1.2$ MHz. The measured EIT line is composed of three transitions to different hyperfine states [$F'=4\rightarrow F''=5,4,3$, see also Fig.~\ref{fig:Experimental config}(b)] whose transition frequencies are spaced over 18.3 MHz. When summing over these transitions with their corresponding oscillator strengths, we find that the measured width $\gamma$ is consistent when accounting for the contributions of two-photon transit time ($2\times \Gamma_\mathrm{tt}=18$ MHz, as calculated and as measured from transient measurements presented in Fig. \ref{fig:transient}), residual longitudinal Doppler width ($2\sigma=1.3$ MHz), laser linewidth (${\sim }0.8$ MHz), and radiative decay rate ($0.66$ MHz).

\section{Conclusions}
We have introduced a new platform to explore light-matter interactions employing a nanofiber-guided mode with a super-extended evanescent field, characterized by low transit-time broadening. This unique mode is reached through adiabatic following of the fundamental mode in a single-mode fiber tapered down to a quarter of a wavelength. Our demonstration, with high overall transmission, sustaining a $13 \lambda$ extended mode over several mm, lies near the asymptotic limit on tapered fiber dimensions capable of such wave-guiding. 

Interfaced with atomic vapor, the system balances high local intensity with an extended coherence time. This balance enables saturation of transmission at optical powers that are on par with much tighter confined modes, and at ultra-low power levels with $\lesssim$ 2 photons present in the interaction region at any given time. We further observe coherent, spectrally-narrow, two-photon resonances, owing to suppression of transit-time broadening. Applying fast in-fiber modulation of the input signal, we observe intricate dynamics of nanofiber-vapor coupled system composed of fast modulation of the optical dipole followed by relatively long relaxation times, which are one order of magnitude longer than previously measured for a thermal vapor-waveguide interface. 
%Having characterized the properties of this unique mode, we intend to use it to explore quantum non-linear phenomena (QNLO) using the Rydberg blockade mechanism. 

The super-extended evanescent mode combines several features which make it particularly well suited for photon-photon interactions and quantum non-linear optics with Rydberg atoms \cite{Peyronel2012}. Firstly, more than 99\% of the guided energy resides outside the material core, as opposed to $\lesssim 50$\% in other evanescent field platforms. In addition, the effective field diameter MFD = 10$~\mu $m is suitable for confining photons to below the Rydberg blockade distance. The proximity of Rydberg atoms to dielectric surfaces induces inhomogeneous Van-der-Waals (VdW) interactions which have been studied, for example, in hollow-core fiber experiments \cite{Epple2014a}. Due to a favorable ratio of mode volume to fiber surface, we expect this detrimental effect to be suppressed in our extremely tapered fiber. Indeed, Rydberg excitations in a cold atomic cloud near a standard nanofiber were recently observed \cite{Rajasree2020}. In addition to VdW and other dipolar interactions, static charges generating stray electric fields pose a major challenge to interfacing a waveguide with Rydberg atoms due to the strong polarizability of the latter. Controlling and removing single charges may therefore be necessary as part of generating Rydberg excitations in our system. Such control and expulsion of charge down to the single electron level was already achieved in several other platforms \cite{Frimmer2017,Moore2014}.
We conclude that new and surprising capabilities can emerge from interfacing the platform presented here with various atomic ensembles for waveguide-based quantum optics and sensing applications.  
%Force and acceleration sensing with optically levitated nanogram masses at microkelvin temperatures, Search for millicharged particles using optically levitated microspheres.,Controlling the net charge on a nanoparticle optically levitated in vacuum .

\section*{Acknowledgments}
We thank Liron Stern and Uriel Levy for fruitful discussions. We acknowledge financial support by the Israel Science Foundation and ICORE, the European Research Council starting investigator grant Q-PHOTONICS 678674, the Minerva Foundation with funding from the Federal German Ministry for Education and Research, the Laboratory in Memory of Leon and Blacky Broder, and the Alexander von Humboldt Foundation (Alexander von Humboldt-Professorship).

\section*{Methods}

% description of the experimental system
The probe laser is a Vescent distributed bragg reflector laser, which is frequency-offset locked to a stable reference laser locked to a cavity. The control laser is a Toptica external-cavity diode laser, which is side-of-fringe locked to an EIT signal obtained in a vapor cell. Both lasers are inserted into the system via fiber-coupled high bandwidth EOMs (iXblue NIR-MX800-LN-10). The EOM modulation signal is produced by a SRS DG645 delay generator, equipped with a SRD1 fast-risetime module.

The probe signal after exiting the fiber is picked off and filtered, by polarization and wavelength, and measured by an avalanche photo-diode or by a single-photon counting module (SPCM).  To measure the tapered fiber transmission at the intensity level of single photons and with high temporal resolution, we use an Excelitas NIR-14-FC SPCM whose output is fed to a Fast ComTec MCS6 time tagger.  
% fiber fabrication

We use a Fibercore SM800 fiber, which is tapered gradually along 3.3 cm. The tapering profile consists of three stages, a steep linear taper along 4 mm (taper angle $\simeq$ 5 mrad), a moderate linear taper along 18 mm (taper angle $\simeq$ 2 mrad), and an exponential taper along the remaining  11 mm. The nanofiber waist is 5 mm long. 

\bibliography{library_20200812}
%\end{multicols}
\end{document}